\documentclass[aps,twocolumn,prl,superscriptaddress,showpacs]{revtex4}
\usepackage{epsfig, amssymb}

\begin{document}
\advance\hoffset by  -4mm

\newcommand{\de}{\Delta E}
\newcommand{\mbc}{M_{\rm bc}}
\newcommand{\bb}{B{\bar B}}
\newcommand{\qq}{q{\bar q}}
\newcommand{\ks}{K^0_S}
\newcommand{\kstar}{\bar{K}^{*0}}
\newcommand{\kpi}{K^+\pi^-}
\newcommand{\kk}{K^+ K^-}
\newcommand{\kpipin}{\kpi\pi^0}
\newcommand{\kpipipi}{\kpi\pi^-\pi^+}
\newcommand{\dkpi}{\bar{D}^0\to\kpi}
\newcommand{\dkpipin}{\bar{D}^0\to\kpipin}
\newcommand{\dkpipipi}{\bar{D}^0\to\kpipipi}
\newcommand{\phipi}{\phi\pi^+}
\newcommand{\phik}{\phi K^-}
\newcommand{\kstark}{\kstar K^+}
\newcommand{\ksk}{K_S^0 K^+}
\newcommand{\ds}{D_s}
\newcommand{\dsp}{D_s^+}
\newcommand{\dsst}{D_s^*}
\newcommand{\dsstp}{D_s^{*+}}
\newcommand{\dsj}{D_{sJ}}
\newcommand{\dsjp}{D_{sJ}^+}
\newcommand{\dsphipi}{\ds\to\phipi}
\newcommand{\dskstark}{\ds\to\kstark}
\newcommand{\dsksk}{\ds\to\ksk}
\newcommand{\br}{{\cal B}}

\title{\Large \rm Observation of the $\dsj(2317)$ and $\dsj(2457)$
in $B$ decays}
\affiliation{Budker Institute of Nuclear Physics, Novosibirsk}
\affiliation{Chiba University, Chiba}
\affiliation{University of Cincinnati, Cincinnati, Ohio 45221}
\affiliation{University of Frankfurt, Frankfurt}
\affiliation{Gyeongsang National University, Chinju}
\affiliation{University of Hawaii, Honolulu, Hawaii 96822}
\affiliation{High Energy Accelerator Research Organization (KEK), Tsukuba}
\affiliation{Hiroshima Institute of Technology, Hiroshima}
\affiliation{Institute of High Energy Physics, Chinese Academy of Sciences, Beijing}
\affiliation{Institute of High Energy Physics, Vienna}
\affiliation{Institute for Theoretical and Experimental Physics, Moscow}
\affiliation{J. Stefan Institute, Ljubljana}
\affiliation{Kanagawa University, Yokohama}
\affiliation{Korea University, Seoul}
\affiliation{Kyungpook National University, Taegu}
\affiliation{Institut de Physique des Hautes \'Energies, Universit\'e de Lausanne, Lausanne}
\affiliation{University of Ljubljana, Ljubljana}
\affiliation{University of Maribor, Maribor}
\affiliation{University of Melbourne, Victoria}
\affiliation{Nagoya University, Nagoya}
\affiliation{Nara Women's University, Nara}
\affiliation{National Kaohsiung Normal University, Kaohsiung}
\affiliation{National Lien-Ho Institute of Technology, Miao Li}
\affiliation{Department of Physics, National Taiwan University, Taipei}
\affiliation{H. Niewodniczanski Institute of Nuclear Physics, Krakow}
\affiliation{Nihon Dental College, Niigata}
\affiliation{Niigata University, Niigata}
\affiliation{Osaka City University, Osaka}
\affiliation{Panjab University, Chandigarh}
\affiliation{Princeton University, Princeton, New Jersey 08545}
\affiliation{RIKEN BNL Research Center, Upton, New York 11973}
\affiliation{University of Science and Technology of China, Hefei}
\affiliation{Seoul National University, Seoul}
\affiliation{Sungkyunkwan University, Suwon}
\affiliation{University of Sydney, Sydney NSW}
\affiliation{Tata Institute of Fundamental Research, Bombay}
\affiliation{Toho University, Funabashi}
\affiliation{Tohoku Gakuin University, Tagajo}
\affiliation{Tohoku University, Sendai}
\affiliation{Department of Physics, University of Tokyo, Tokyo}
\affiliation{Tokyo Institute of Technology, Tokyo}
\affiliation{Tokyo Metropolitan University, Tokyo}
\affiliation{Tokyo University of Agriculture and Technology, Tokyo}
\affiliation{Toyama National College of Maritime Technology, Toyama}
\affiliation{University of Tsukuba, Tsukuba}
\affiliation{Utkal University, Bhubaneswer}
\affiliation{Virginia Polytechnic Institute and State University, Blacksburg, Virginia 24061}
\affiliation{Yokkaichi University, Yokkaichi}
\affiliation{Yonsei University, Seoul}
 \author{P.~Krokovny}\affiliation{Budker Institute of Nuclear Physics, Novosibirsk} 
  \author{K.~Abe}\affiliation{High Energy Accelerator Research Organization (KEK), Tsukuba} 
  \author{K.~Abe}\affiliation{Tohoku Gakuin University, Tagajo} 
  \author{T.~Abe}\affiliation{High Energy Accelerator Research Organization (KEK), Tsukuba} 
  \author{I.~Adachi}\affiliation{High Energy Accelerator Research Organization (KEK), Tsukuba} 
  \author{H.~Aihara}\affiliation{Department of Physics, University of Tokyo, Tokyo} 
  \author{K.~Akai}\affiliation{High Energy Accelerator Research Organization (KEK), Tsukuba} 
  \author{M.~Akatsu}\affiliation{Nagoya University, Nagoya} 
  \author{M.~Akemoto}\affiliation{High Energy Accelerator Research Organization (KEK), Tsukuba} 
  \author{Y.~Asano}\affiliation{University of Tsukuba, Tsukuba} 
% ...
\collaboration{The Belle Collaboration}

\begin{abstract}
We report the first observation of the $B\to\bar{D}\dsj(2317)$ and 
$B\to\bar{D}\dsj(2457)$ decays based on $123.8\times 10^6$ $\bb$ events
collected with the Belle detector at KEKB.
We observe the $\dsj(2317)$ decay to $\ds\pi^0$ and $\dsj(2457)$ decay to
the $\dsst\pi^0$ and $\ds\gamma$ final states. We also set 90\% CL
upper limits for the decays $\dsj(2317)\to\dsst\gamma$,
$\dsj(2457)\to\dsst\gamma$, $\dsj(2457)\to\ds\pi^0$ and
$\dsj(2457)\to\ds\pi^+\pi^-$.
\end{abstract}
\pacs{13.25.Hw, 14.40.Lb}
\maketitle

Recently a new $\ds\pi^0$ resonance with a mass of 2317~MeV$/c^2$ and a
very narrow width was observed by the BaBar 
collaboration~\cite{babar_dspi0}. A natural interpretation
is that this is a $P$-wave $c\bar{s}$ quark state that is below
the $D K$ threshold, which accounts for the small 
width~\cite{bardeen}. This interpretation is supported by the
observation of a $\dsst\pi^0$ resonance~\cite{footnote1}
by the CLEO collaboration~\cite{cleo_dspi0} and Belle 
collaboration~\cite{belle_dspi0}.
All groups observe these states in inclusive $e^+e^-$ processes.
The mass difference between the two observed states is consistent with
the expected hyperfine splitting of the $P$-wave $\ds$ 
meson doublet with total light-quark angular momentum
$j=1/2$~\cite{bardeen}. However, the masses of these states are 
considerably below potential model expectations~\cite{bartelt}, and
are nearly the same as those of the corresponding $c\bar{u}$ states
recently measured by Belle~\cite{kuzmin}.
The low mass values have caused speculation that these
states may be more exotic than a simple $q\bar{q}$ meson 
system~\cite{cahn,lipkin,beveren,hou,fazio,godfrey}.
To clarify the nature of these states, it is necessary to
determine their quantum numbers and decay branching fractions,
particularly those for radiative decays. In this context it is
useful to search for these states, which we refer to as
$\dsj$, in exclusive $B$ meson decay processes.
  
We search for decays of the type $B\to\bar{D}\dsj$, which are expected 
to be the dominant exclusive $\dsj$ production mechanism in $B$ decays.
Because of the known properties of the parent $B$ meson, angular 
analyses of these decays can unambiguously determine the $\dsj$ 
quantum numbers. Moreover, since QCD sum rules in HQET predict
that $P$-wave mesons with $j=1/2$ should be more readily produced in $B$
decays than mesons with $j=3/2$~\cite{yao}, the observation of 
$B\to\bar{D}\dsj$ would provide additional support for the $P$-wave 
nature of these states as well as serving as a check of these predictions.

In this Letter we report on a search for the $B\to\bar{D}\dsj(2317)$ and
$B\to\bar{D}\dsj(2457)$ decays based on a sample of $123.8\times 10^6$ 
$\bb$ pairs produced at the KEKB asymmetric energy $e^+e^-$ 
collider~\cite{KEKB}. The inclusion of charge conjugate states is 
implicit throughout this report.

The Belle detector has been described elsewhere~\cite{NIM}.
Charged tracks are selected with a set of requirements based on the
average hit residual and impact parameter relative to the
interaction point (IP).
The transverse momentum of at least
0.05~GeV$/c$ is required for each track
in order to reduce the combinatorial background. 

For charged particle identification (PID), the combined information
from specific ionization in the central drift chamber ($dE/dx$),
time-of-flight scintillation counters and aerogel \v{C}erenkov
counters is used.
Charged kaons are selected with PID criteria that have
an efficiency of 88\%, a pion misidentification probability of 8\%,
and negligible contamination from protons.
All charged tracks with PID responses consistent with a pion
hypothesis that are not positively identified as electrons are 
considered as pion candidates.

Neutral kaons are reconstructed via the decay $K_S^0\to\pi^+\pi^-$
with no PID requirements for the daughter pions.
The two-pion invariant mass is required to be within 9~MeV$/c^2$
($\sim 3\sigma$) of the $K^0$ mass and the displacement of the 
$\pi^+\pi^-$ vertex from the IP in the transverse ($r-\varphi$) plane is 
required to be between 0.2~cm and 20~cm. 
The direction in the $r-\varphi$ plane from the IP to the $\pi^+\pi^-$ 
vertex is required to agree within 0.2 radians with the combined 
momentum of the two pions.

Photon candidates are selected from calorimeter showers not associated
with charged tracks.
An energy deposition of at least 30~MeV and a photon-like shape are 
required for each candidate.
A pair of photons with an invariant mass 
within 12~MeV$/c^2$ ($\sim 2.5\sigma$) of the $\pi^0$ mass is 
considered as a $\pi^0$ candidate.

\begin{figure*}
  \includegraphics[width=0.47\textwidth] {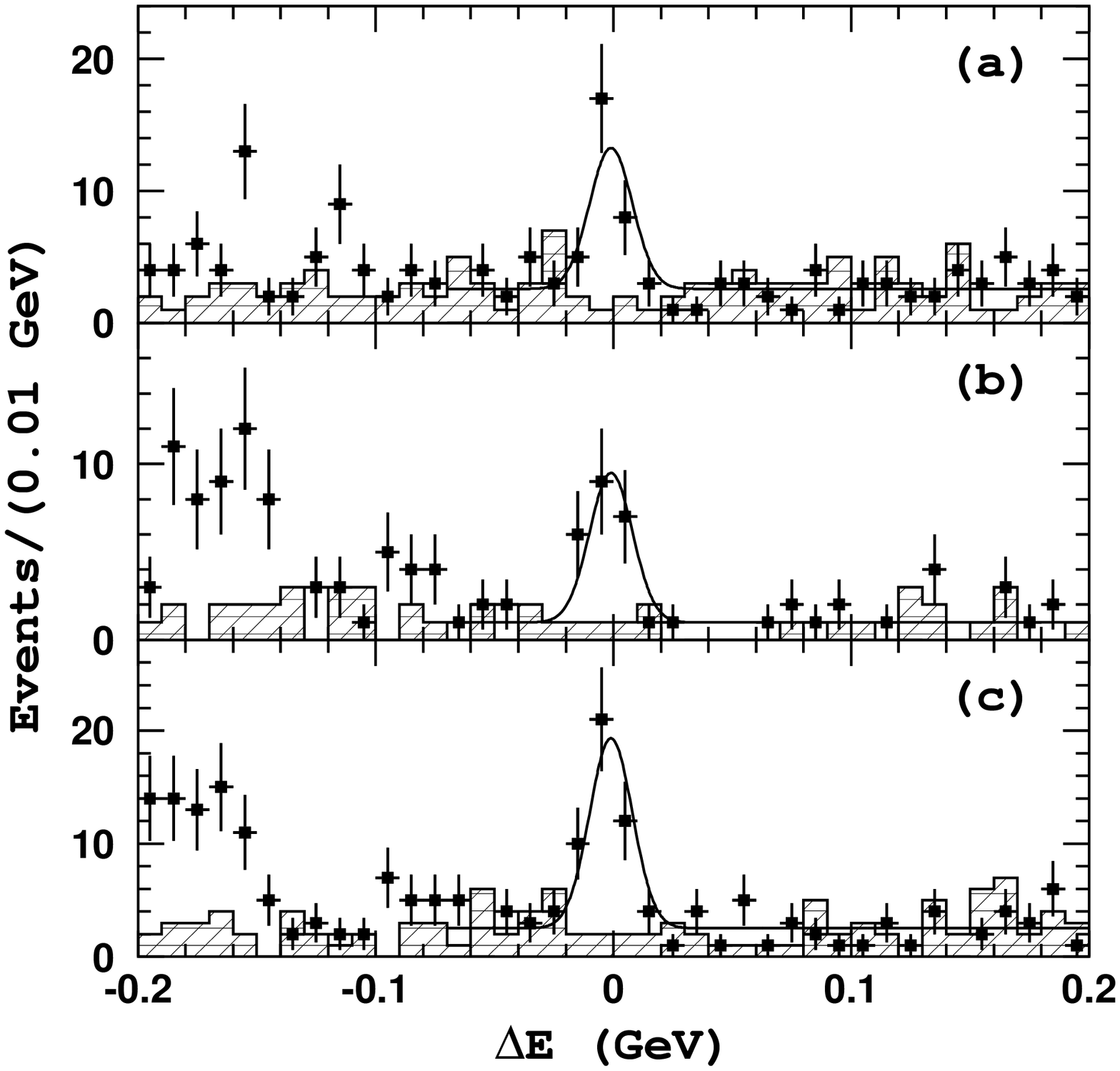} \hfill
  \includegraphics[width=0.47\textwidth] {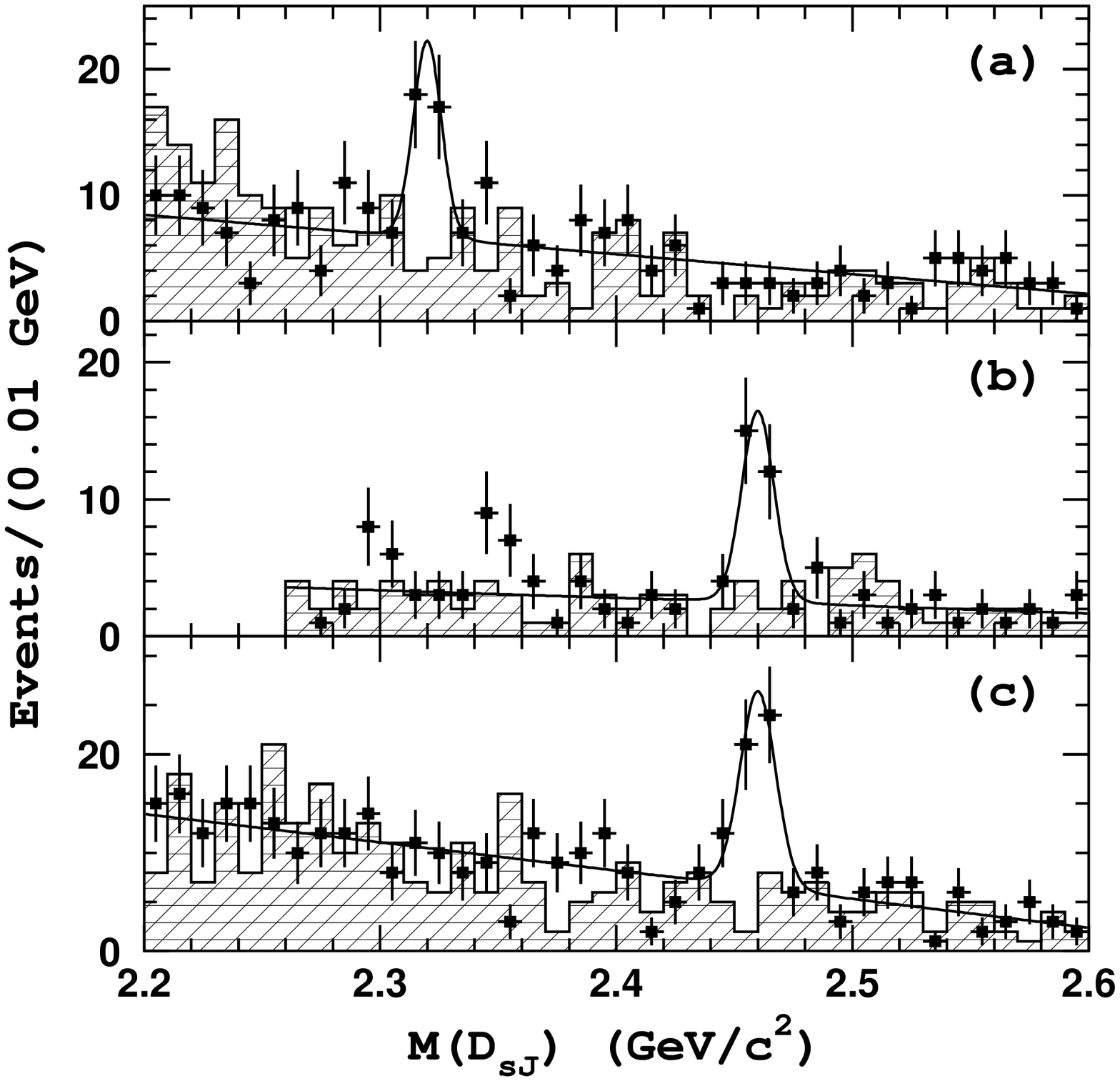}
  \caption{$\de$ (left) and $M(\dsj)$ (right) distributions for the
    $B\to\bar{D}\dsj$ candidates: (a) $\dsj(2317)\to\ds\pi^0$, (b)
    $\dsj(2457)\to\dsst\pi^0$ and (c) $\dsj(2457)\to\ds\gamma$. 
    Points with errors
    represent the experimental data, crosshatched histograms show
    the sidebands and curves are the results of the fits.}
  \label{de_all}
\end{figure*}

We reconstruct $\bar{D}^0 (D^-)$ mesons in the $\kpi$, $\kpipipi$ and 
$\kpipin$ $(\kpi\pi^-)$ decay channels and require the invariant mass 
to be within 12~MeV$/c^2$ ($1.5\sigma$ for $\kpipin$ and 
$2.5\sigma$ for other modes) of the $\bar{D}^0 (D^-)$ mass.
For the $\pi^0$ from the $\dkpipin$ decay, we require that
the $\pi^0$ momentum in the $\Upsilon(4S)$ center-of-mass (CM) frame
be greater than 0.4~GeV$/c$ in order to reduce combinatorial backgrounds.
We reconstruct $\dsp$ mesons in the $\phipi$, $\kstark$ and $\ksk$
decay channels.
$\phi$ mesons are reconstructed from $\kk$ pairs with an
invariant mass within 10~MeV$/c^2$ ($2.5\Gamma$) of the
$\phi$ mass. $\kstar$ mesons are reconstructed from $K^-\pi^+$
pairs with an invariant mass within 75~MeV$/c^2$ ($1.5\Gamma$) of the
$\kstar$ mass. After calculating the invariant mass of the
corresponding set of particles, we define the $\dsp$ signal region 
as being within 12~MeV$/c^2$ ($\sim 2.5\sigma$) of the $\ds$ 
mass. $\dsst$ mesons are reconstructed in the $\dsst\to\ds\gamma$
decay channel. The mass difference between $\dsst$ and $\ds$
candidates is required to be within 8~MeV$/c^2$ of its
nominal value ($\sim 2.5\sigma$).
The $\dsj$ candidates are reconstructed from $D_s^{(*)}$ mesons and 
a $\pi^0$, $\gamma$, or $\pi^+\pi^-$ pair. The mass difference 
$M(\dsj)-M(D_s^{(*)})$ is used to select $\dsj$ candidates.
We use central mass values of 2317~MeV$/c^2$ and 2460~MeV$/c^2$
for $\dsj(2317)$ and $\dsj(2457)$ respectively and define signal 
regions within 12~MeV$/c^2$ for the corresponding mass difference.

We combine $\bar{D}$ and $\dsj$ candidates to form $B$ mesons.
Candidate events are identified by their CM energy difference, 
\mbox{$\de=(\sum_iE_i)-E_{\rm beam}$}, and the beam constrained mass, 
$\mbc=\sqrt{E_{\rm beam}^2-(\sum_i\vec{p}_i)^2}$, where $E_{\rm beam}$ 
is the beam energy and $\vec{p}_i$ and $E_i$ are the momenta and 
energies of the decay products of the $B$ meson in the CM frame. 
We select events with $5.272$~GeV$/c^2<\mbc<5.288$~GeV$/c^2$
and $|\de|<0.2$~GeV, and define a $B$ signal region of $|\de|<0.03$~GeV.
In cases with more than one candidate in an event, the one with
$D$ and $D_s^{(*)+}$ masses closest to the nominal values is chosen.
We use a Monte Carlo (MC) simulation to model the response of
the detector and determine the efficiency~\cite{GEANT}.


Variables that characterize the event topology are used to suppress 
background from the two-jet-like $e^+e^-\to\qq$ continuum process.
We require $|\cos\theta_{\rm thr}|<0.80$, where $\theta_{\rm thr}$ is 
the angle between the thrust axis of the $B$ candidate and that of the 
rest of the event; this eliminates 77\% of the continuum 
background while retaining 78\% of the signal events. 
To suppress  combinatorial background we apply a restriction on the
invariant mass of the $D$ meson and the $\pi^0$ or $\gamma$ from 
$\dsj$ decay: $M(D\pi^0)>2.3$~GeV$/c^2$, $M(D\gamma)>2.2$~GeV$/c^2$.

\begin{table*}
\caption{Product branching fractions for $B\to\bar{D}\dsj$ decays.}
\medskip
\label{defit}
\begin{tabular*}{\textwidth}{l@{\extracolsep{\fill}}ccccc}\hline\hline
Decay channel & $\de$ yield & $M(\dsj)$ yield & Efficiency, $10^{-4}$ & 
${\cal B}$, $10^{-4}$ & Significance\\\hline
$B^+\to\bar{D}^0 \dsjp(2317)~[\dsp\pi^0]$, &
$13.7^{+5.1}_{-4.5}$ & $13.4^{+6.2}_{-5.4}$ & 1.36 & 
$8.1^{+3.0}_{-2.7}\pm 2.4$ & $5.0\sigma$\\
$B^0\to D^- \dsjp(2317)~[\dsp\pi^0]$ &
$10.3^{+3.9}_{-3.1}$ & $10.8^{+4.2}_{-3.6}$ & 0.97 &
$8.6^{+3.3}_{-2.6}\pm 2.6$ & $6.1\sigma$\\

$B^+\to\bar{D}^0 \dsjp(2317)~[\dsstp\gamma]$ &
$3.4^{+2.8}_{-2.2}$ & $2.1^{+4.1}_{-3.4}$ & 1.08 & 
$2.5^{+2.1}_{-1.6}(<7.6)$ & ---\\
$B^0\to D^- \dsjp(2317)~[\dsstp\gamma]$ &
$2.3^{+2.5}_{-1.9}$ & $1.6^{+2.4}_{-1.9}$ & 0.69 & 
$2.7^{+2.9}_{-2.2}(<9.5)$ & ---\\

$B^+\to\bar{D}^0 \dsjp(2457)~[\dsstp\pi^0]$ &
$7.2^{+3.7}_{-3.0}$ & $8.9^{+4.0}_{-3.3}$ & 0.49 & 
$11.9^{+6.1}_{-4.9}\pm 3.6$ & $2.9\sigma$\\
$B^0\to D^- \dsjp(2457)~[\dsstp\pi^0]$ &
$11.8^{+3.8}_{-3.2}$ & $14.9^{+4.4}_{-3.9}$ & 0.42 & 
$22.7^{+7.3}_{-6.2}\pm 6.8$ & $6.5\sigma$\\

$B^+\to\bar{D}^0 \dsjp(2457)~[\dsp\gamma]$ &
$19.1^{+5.6}_{-5.0}$ & $20.2^{+7.2}_{-6.9}$ & 2.75 & 
$5.6^{+1.6}_{-1.5}\pm 1.7$ & $5.0\sigma$\\
$B^0\to D^- \dsjp(2457)~[\dsp\gamma]$ &
$18.5^{+5.0}_{-4.3}$ & $19.6^{+5.6}_{-4.9}$ & 1.83 & 
$8.2^{+2.2}_{-1.9}\pm 2.5$ & $6.5\sigma$\\

$B^+\to\bar{D}^0 \dsjp(2457)~[\dsstp\gamma]$ &
$4.4^{+3.8}_{-3.3}$ & $8.2^{+4.0}_{-3.4}$ & 1.15 & 
$3.1^{+2.7}_{-2.3}(<9.8)$ & ---\\
$B^0\to D^- \dsjp(2457)~[\dsstp\gamma]$ &
$1.1^{+1.8}_{-1.2}$ & $0.2^{+1.8}_{-1.2}$ & 0.71 & 
$1.3^{+2.0}_{-1.4}(<6.0)$ & ---\\

$B^+\to\bar{D}^0 \dsjp(2457)~[\dsp\pi^+\pi^-]$ &
$<4.0$ & $-2.2^{+2.0}_{-1.6}$ & 1.89 & $<2.2$ & ---\\
$B^0\to D^- \dsjp(2457)~[\dsp\pi^+\pi^-]$ & 
$<2.5$ & $-1.2^{+2.7}_{-2.0}$ & 1.35 & $<2.0$ & ---\\

$B^+\to\bar{D}^0 \dsjp(2457)~[\dsp\pi^0]$ & $<2.4$ & 
$1.0^{+2.7}_{-2.0}$ & 0.94 & $<2.7$ & ---\\
$B^0\to D^- \dsjp(2457)~[\dsp\pi^0]$ & $<2.4$ & 
$0.3^{+1.8}_{-1.2}$ & 0.68 & $<3.6$ & ---\\
\hline\hline
\end{tabular*}
\end{table*}

\begin{table}
\caption{Combined fit results.}
\medskip
\label{simulfit}
\begin{tabular}{lcc}\hline\hline
Decay channel & ${\cal B}$, $10^{-4}$ & Significance\\\hline
$B\to\bar{D} \dsj(2317)~[\ds\pi^0]$ & 
$8.5^{+2.1}_{-1.9}\pm 2.6$ & $6.1\sigma$\\
$B\to\bar{D} \dsj(2317)~[\dsst\gamma]$ & 
$2.5^{+2.0}_{-1.8}(<7.5)$ & $1.8\sigma$\\

$B\to\bar{D}\dsj(2457)~[\dsst\pi^0]$ & 
$17.8^{+4.5}_{-3.9}\pm 5.3$ & $6.4\sigma$\\

$B\to\bar{D}\dsj(2457)~[\ds\gamma]$ &
$6.7^{+1.3}_{-1.2}\pm 2.0$ & $7.4\sigma$\\

$B\to\bar{D}\dsj(2457)~[\dsst\gamma]$ &
$2.7^{+1.8}_{-1.5}(<7.3)$ & $2.1\sigma$\\

$B\to\bar{D}\dsj(2457)~[\ds\pi^+\pi^-]$ & $<1.6$ & ---\\

$B\to\bar{D}\dsj(2457)~[\ds\pi^0]$ & $<1.8$ & ---\\\hline\hline
\end{tabular}
\end{table}

The $\de$ and $\dsj$ candidate's invariant mass ($M(\dsj)$) 
distributions for $B\to\bar{D}\dsj$ candidates are presented in 
Fig.~\ref{de_all}, where all $\bar{D}^0$ and $D^-$ decay modes are
combined. Each distribution is the projection of the signal region of 
the other parameter; distributions for events in the $M(\dsj)$ and 
$\de$ sidebands are shown as crosshatched histograms.
Clear signals are observed for the $D\dsj(2317)[\ds\pi^0]$ and 
$D\dsj(2457)[\dsst\pi^0, \ds\gamma]$ final states.
The measured masses for the $\dsj(2317)$ and $\dsj(2457)$ are
$(2319.8\pm 2.1\pm 2.0)$~MeV$/c^2$ and $(2459.2\pm 1.6\pm
2.0)$~MeV$/c^2$ respectively. The fitted widths are consistent with 
those expected for $\dsj$ mesons of zero intrinsic width.
The systematic error in the $\dsj$ mass is expected to come from 
the photon energy scale.

\begin{figure}
  \includegraphics[width=0.4\textwidth] {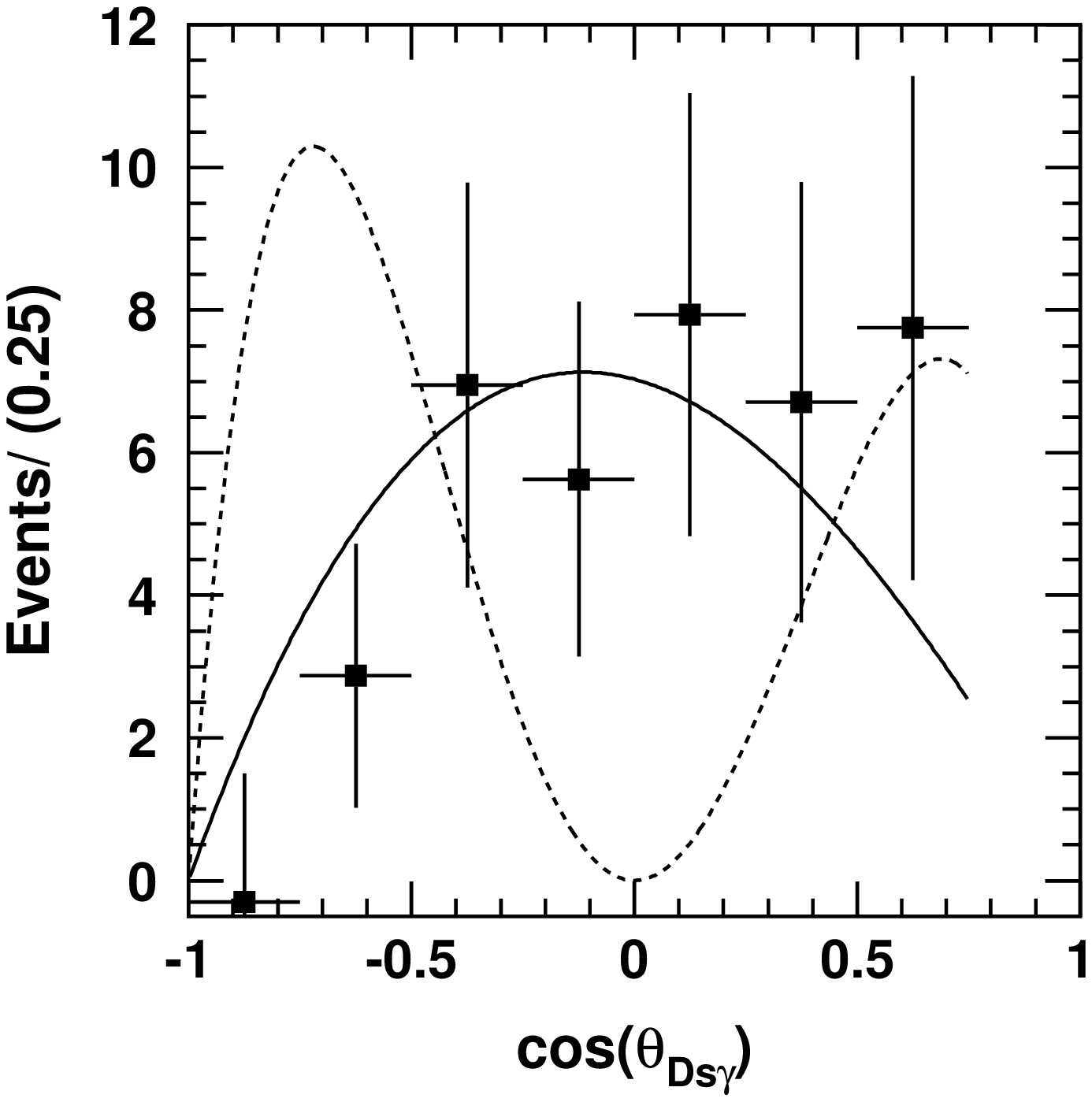}
  \caption{The $\dsj(2457)\to\ds\gamma$ helicity distribution.
   The points with error bars are the results of fits to the $\de$ 
   spectra for experimental events. 
   Solid and dashed curves are MC predictions for the $J=1$ and $J=2$
   hypotheses, respectively. The highest bin has 
   no events because of the cut on the $D\gamma$ invariant mass.} 
  \label{check}
\end{figure}

We also study the helicity distribution for the
$\dsj(2457)\to\ds\gamma$ decay.
The helicity angle $\theta_{\ds\gamma}$ is defined as the angle
between the $\dsj(2457)$ momentum in the $B$ meson rest frame and the 
$\ds$ momentum in the $\dsj(2457)$ rest frame.
The $\theta_{\ds\gamma}$ distribution in the data (Fig.~\ref{check}) is 
consistent with MC expectations for the $J=1$ hypothesis for the 
$\dsj(2457)$ ($\chi^2/$n.d.f$=5/6$), and contradicts the $J=2$ hypothesis 
($\chi^2/$n.d.f.$=44/6$). The $J=0$ hypothesis is already ruled out by 
the conservation of angular momentum and parity in 
$\dsj(2457)\to\ds\gamma$.

For each decay channel, the $\de$ distribution is fitted with a 
Gaussian signal and a linear background function. The Gaussian 
mean value and width are fixed to the values from a MC simulation of
signal events. The region $\de<-0.07$~GeV is excluded from the fit 
to avoid contributions from other $B$ decays of the type
$B\to\bar{D}\dsj X$ where $X$ denotes an additional particle 
that is not reconstructed.
The $M(\dsj)$ distribution is fitted by the sum of a 
Gaussian
for the signal, and a linear function for the background. 
The Gaussian width is fixed to 
the value found in the MC (6--7~MeV$/c^2$ depending on the decay mode).
The fit results are given in Table~\ref{defit}, where the listed
efficiencies include intermediate branching fractions.  
We use the $\de$ distribution to calculate the branching fractions.
The statistical significance of the signal quoted in Table~\ref{defit} 
is defined as
$\sqrt{-2\ln({\cal L}_0/{\cal L}_{max})}$, where ${\cal L}_{max}$ and
${\cal L}_0$ denote the maximum likelihood with the nominal
and with zero signal yield, respectively.

The results of combined fits of $B^+\to\bar{D}^0\dsj^+$ and 
$B^0\to D^-\dsj^+$ modes assuming isospin invariance are shown
in Table~\ref{simulfit}.
The normalization of the background in each sub-mode is allowed to 
float while the signal yields are required to satisfy the constraint
$N_i= N_{\bb}\cdot{\cal B}(B\to\bar{D}\dsj)\cdot\varepsilon_i\, ,$
where the branching fraction ${\cal B}(B\to\bar{D}\dsj)$ is a fit 
parameter;
$N_{\bb}$ is the number of $\bb$ pairs and $\varepsilon_i$ is the 
efficiency, which includes all intermediate branching fractions.
From the two $B\to\bar{D}\dsj(2457)$ branching fraction measurements, we 
determine the ratio $\br(\dsj(2457)\to\ds\gamma)/
\br(\dsj(2457)\to\dsst\pi^0)=0.38\pm0.11\pm 0.04$.

\begin{figure}
  \includegraphics[width=0.42\textwidth] {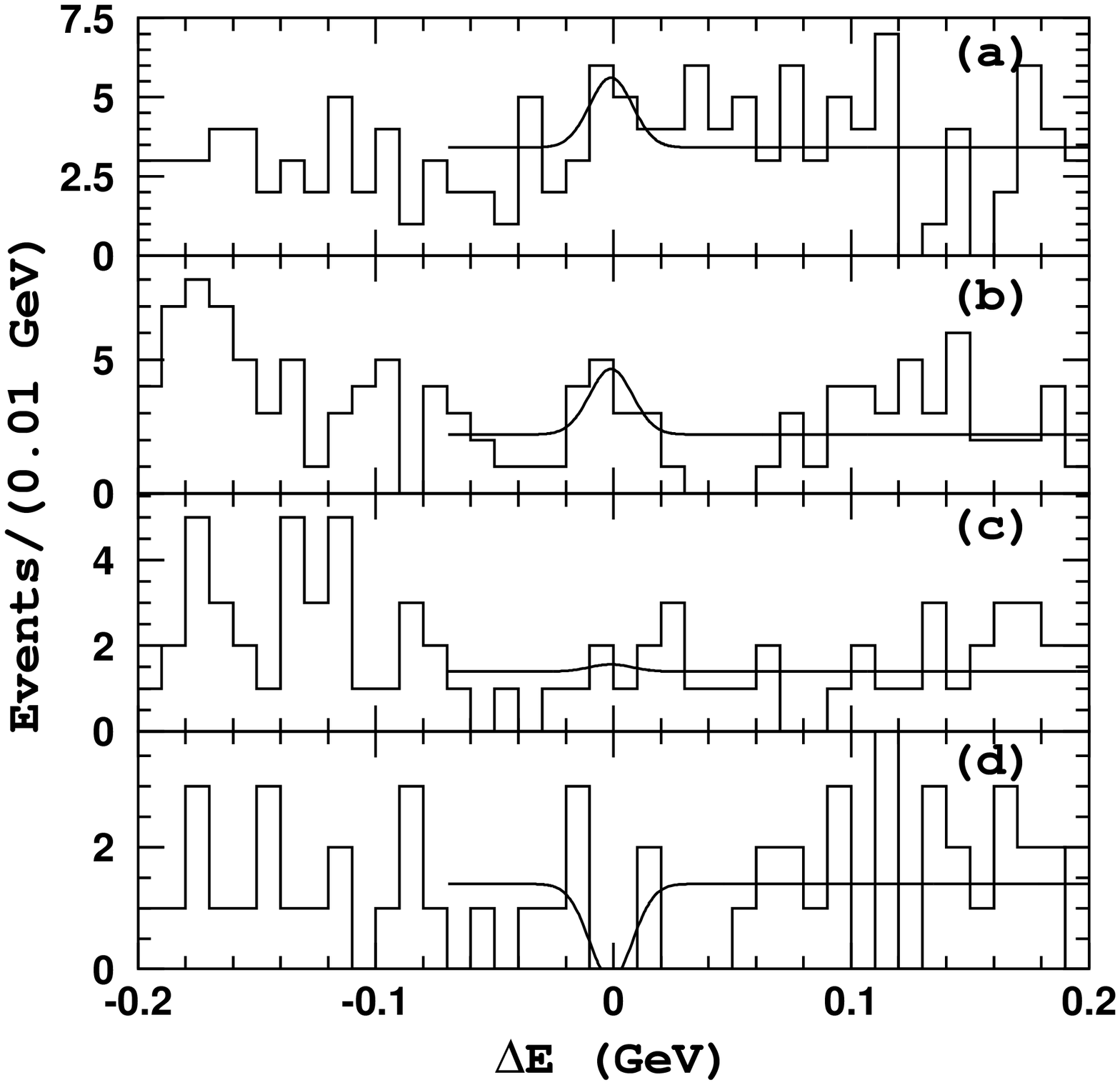}
  \caption{$\de$ distributions for the decay channels with not
    significant signals: (a) $D\dsj(2317)[\dsst\gamma]$, (b)
    $D\dsj(2457)[\dsst\gamma]$, (c) $D\dsj(2457)[\ds\pi^+\pi^-]$, 
    (d) $D\dsj(2457)[\ds\pi^0]$.
    Open histograms represent the experimental data
    and curves show the results of the fits.}
  \label{de_ul}
\end{figure}

The signals for the $B\to\bar{D} \dsj(2317)[\ds\pi^0]$ and
$B\to\bar{D} \dsj(2457)[\dsst\pi^0, \ds\gamma]$ channels have 
greater than $5\sigma$ statistical significance.
Figure~\ref{de_ul} shows the $\de$ distributions for the
other channels, where significant signals are not seen. 
We set 90\% confidence level (CL) upper limits for 
these modes.

We study the possible feed-across between all studied $\dsj$ decay
modes using MC. We also analyze a MC sample of generic $\bb$ events 
corresponding to our data sample. No peaking background is found.
As a check, we apply a similar procedure to decay chains with the 
similar final states: $B\to \bar{D}^{(*)}D_s^{(*)}$.
For each mode, we measure branching fractions that 
are consistent with the world average values~\cite{PDG}.

The following sources of systematic errors are considered:
tracking efficiency (1-2\% per track), kaon identification
efficiency (1\%), $\pi^0$ efficiency (6\%), $K^0_S$ reconstruction 
efficiency (6\%), $D$ branching fraction
uncertainties (2\%-6\%), signal and background shape
parameterization (4\%) and MC statistics (3\%).
The uncertainty in the tracking efficiency is estimated using 
partially reconstructed $D^{*+}\to D^0[K_S^0\pi^+\pi^-]\pi^+$ decays. 
The kaon identification uncertainty is determined from 
$D^{*+}\to D^0[K^-\pi^+]\pi^+$ decays. The $\pi^0$ reconstruction 
uncertainty is obtained using $D^0$ decays to $K^-\pi^+$ and 
$K^-\pi^+\pi^0$.
We assume equal production rates for $B^+B^-$ and $B^0\bar B^0$ pairs 
and do not include the uncertainty related to this assumption in the 
total systematic error. For the calculation of the branching
fractions, the errors in the $\ds$ meson branching fractions are
taken into account. These uncertainties are dominated by the error on
the $\dsphipi$ branching ratio of 25\%~\cite{PDG}. The overall
systematic uncertainty is 30\%.


In summary, we report the first observation of $B\to\bar{D}\dsj(2317)$ 
and $B\to\bar{D}\dsj(2457)$ decays. 
The measured branching fractions with the corresponding statistical
significances are presented in Table~\ref{simulfit}.
The observation of $\dsj(2457)\to\ds\gamma$ decay eliminates the zero
spin of $\dsj(2457)$. The angular analysis of this decay supports
the hypothesis that the $\dsj(2457)$ is a $1^+$ state.

We wish to thank the KEKB accelerator group for the excellent
operation of the KEKB accelerator.
We acknowledge support from the Ministry of Education,
Culture, Sports, Science, and Technology of Japan
and the Japan Society for the Promotion of Science;
the Australian Research Council
and the Australian Department of Education, Science and Training;
the National Science Foundation of China under contract No.~10175071;
the Department of Science and Technology of India;
the BK21 program of the Ministry of Education of Korea
and the CHEP SRC program of the Korea Science and Engineering Foundation;
the Polish State Committee for Scientific Research
under contract No.~2P03B 01324;
the Ministry of Science and Technology of the Russian Federation;
the Ministry of Education, Science and Sport of the Republic of Slovenia;
the National Science Council and the Ministry of Education of Taiwan;
and the U.S.\ Department of Energy.


\begin{thebibliography}{99}
\bibitem{footnote1}{In the heavy $c$-quark mass limit, one expects two 
  doublets of $c\bar{s}$ states with quantum 
  numbers $J^P=0^+$, $1^+$ and $1^+$, $2^+$. The second one has 
  been observed in $D^{(*)}K$ decays.}
\end{thebibliography}
\end{document}